\documentclass[a4paper]{jpconf}
\usepackage{graphicx}
\usepackage{amsmath}
\usepackage{amssymb}

\begin{document}
\title{Breaking stress of Coulomb crystals in the neutron star crust}

\author{A A Kozhberov}

\address{Ioffe Institute, Politekhnicheskaya 26,
St. Petersburg, 194021, Russia}

\ead{kozhberov@gmail.com}

\begin{abstract}
It is generally accepted that the Coulomb crystal model can be used to describe matter in the neutron star crust. In \cite{KY20} we study the properties of deformed Coulomb crystals and how their stability depends on the polarization of the electron background; the breaking stress in the crust $\sigma_{\max}$ at zero temperature was calculated based on the analysis of the electrostatic energy and the phonon spectrum of the Coulomb crystal. In this paper, I briefly discuss the influence of zero-point and thermal contributions on $\sigma_{\max}$.
\end{abstract}

\section{Introduction}

Neutron stars are born in the final stages of stellar evolution. The structure and composition of matter in neutron star has been widely discussed, but remain not completely clear, especially for the core (e.g., \cite{ST83,HPY07}). The structure of the neutron star crust ($\rho \lesssim 1.5 \times 10^{14}$ g cm$^{-3}$)
seems to be more clear (e.g., \cite{ST83,HPY07,CH08,CF16,CH17}). At $\rho >10^4$ g cm$^{-3}$ the crust consists of an electron gas and fully ionized atomic nuclei. The outer layer of the crust is composed of iron nuclei ($^{56}$Fe). With increasing density ($\rho \gtrsim 8 \times 10^6$  g cm$^{-3}$) nuclei become progressively more neutron-rich. At $\rho\approx (4-6) \times 10^{11}$ g cm$^{-3}$ free neutrons appear in matter but they have a rather weak effect on the dynamic behavior of electrons and ions.

The interaction between $i$th and $j$th point-like ion with charge $Ze$ and mass $M$ in the neutron star crust can be described by the screened Coulomb potential
\begin{equation}
U(r_{ij})=Z^2 e^2 \frac{\exp(-\kappa r_{ij})}{r_{ij}}~,
\label{Yuk}
\end{equation}
where $r_{ij}$ is the distance between two ions; $\kappa\equiv \sqrt{4\pi e^2\partial n_e/\partial\mu_e}$ is the electron screening parameter; $n_e$, $\mu_e$, and  $e$ are the electron number density, chemical potential, and charge, respectively. It is more convenient to use the dimensionless screening parameter $\kappa a$, where $a\equiv(4\pi n/3)^{-1/3}$ is the ion sphere radius, $n=n_e/Z$ is the number density of ions. For degenerate electrons
\begin{equation}
   \kappa a \approx 0.1850\,Z^{1/3}\,\frac{(1+x^2)^{1/4}}{x^{1/2}},
\label{e:sTF}   
\end{equation}
where $x \approx 0.01 (\rho Z/A)^{1/3}$ is the electron relativity parameter, $A$ is the mass number of ions, $\rho$  is the density (in g cm$^{-3}$ units).

If $\kappa a \lesssim 1$, the ions form a crystal, which is usually called a Coulomb or Yukawa, at $\Gamma\equiv Z^2 e^2/(aT) \sim 200$ \cite{HFD97}. The distance between two ions in a Coulomb crystal is $r_{ij}=|\textbf{R}_i-\textbf{R}_j+\textbf{u}_i-\textbf{u}_j|$, where $\textbf{R}_i$ is the equilibrium position of the $i$th ion in the crystal, and $\textbf{u}_i$ is its displacement. At $\Gamma \gg 200$, the motions of ions in the crystal can be considered as small phonon oscillations around equilibrium positions.

The lattice is treated as unstable if one or several of the squared frequencies of the phonon modes are negative at some wave vector. For a body-centered cubic (bcc) lattice the critical value $\kappa a=4.76$ was calculated in \cite{RKG88,K18}. For $^{56}$Fe ions and degenerate electrons this value gives $\rho \approx 5$ g cm$^{-3}$, for which the Coulomb crystal model is inapplicable. In the most part of the neutron star crust $\kappa a \lesssim 1$. Note that I am using a bcc lattice because this lattice possesses the lowest electrostatic energy at $\kappa = 0$ (e.g., \cite{CF16}).

As it shown in \cite{KY20} the deformed lattice becomes unstable at $\kappa a$ less than 4.76 and it significantly depends on the direction of deformation. Here I consider only one deformation of the bcc lattice, which translates the vector $\textbf{R}_i$ as
\begin{equation}
a_{\rm l}(n_1,n_2,n_3)\to
a_{\rm l}\left(n_1+\frac{\epsilon}{2}n_2,
n_2+\frac{\epsilon}{2}n_1,\frac{n_3}{1-\epsilon^2/4}\right),
\label{e:trans1}
\end{equation}
where $a_{\rm l}$ is a lattice constant, $\epsilon$ being a deformation parameter, and $n_1$, $n_2$, and $n_3$ are arbitrary integers. In addition to \cite{KY20} this tensile-shear deformation was considered in \cite{CH10,CH12}, where deformed crystals in neutron stars were studied via molecular dynamic (MD) simulations. 

Our investigations of the phonon spectrum show that the bcc lattice remains stable at $\epsilon \leq \epsilon_{\max}$. The dependence of the maximal deformation parameter $\epsilon_{\max}$ on $\kappa a$ was discussed in \cite{KY20}. For a crystal with a uniform electron background, it is equal 0.11090. Here I will consider a lattice with $\kappa a=4/7$, for it $\epsilon_{\max}=0.10895$.

\section{The breaking stress}
When a neutron star evolves, its crust undergoes various deformations (e.g.,\cite{O05,CH08,BK17,FHL18,G18}). For magnetars, these deformations are primarily associated with the magnetic field (e.g.,\cite{BL14,L16,LL16}); for pulsars they are thought to be connected with glitches (e.g.,\cite{PFH14}). Investigations of deformed lattices are important for understanding different processes in the neutron stars interior.

In most of the previous studies, deformed Coulomb crystals were used to calculate elastic properties and effective shear modulus $\mu$ (e.g.,\cite{OI90,S91,HH08,HK09,H12}). Less attention was paid to the stability of deformed crystals in the neutron star crust and the breaking (maximum) stress.
Analytically it has been done in \cite{BK17,BC18} for a bcc lattice with a uniform electron background. Deformations of lattices with $\kappa a > 0$ have been studied in \cite{CH10,HH12,CH12} via Monte-Carlo simulations and for a restricted number of $\kappa a$ values. The first analytical investigation of the stability of a deformed bcc Coulomb crystal with $\kappa a > 0$ was presented in \cite{KY20}.

The effective stress for any $\epsilon$ can be calculated as
\begin{equation}
   \sigma(\epsilon)=\frac{\partial E_\text{int}/V}{\partial \epsilon},
\label{sig}
\end{equation} 
while the breaking stress is  
\begin{equation}
  \sigma_{\max} \equiv \sigma(\epsilon_{\max}),
\end{equation}
where $E_\text{int}$ is the internal energy of the crystal. 

At zero temperature in the harmonic approximation, the internal energy is a sum of electrostatic $U_{\rm M}$ and zero-point vibration $E_0$ energies. It can be written as
\begin{equation}
U_{\rm M}+E_0\equiv N\frac{Z^2e^{2}}{a}\zeta+1.5N\langle\omega\rangle 
=N\omega_{\rm p}\left(\Gamma_{\rm p}\zeta+1.5 u_1\right)~, \label{f_0}
\end{equation}
where $\zeta$ is the Madelung constant \cite{B02},
$\langle\omega\rangle$ is the phonon frequency averaged over the first Brillion zone,
$u_1 \equiv \langle\omega/\omega_{\rm p}\rangle$ is the first momentum of the phonon spectrum,
$\omega_{\rm p}\equiv\sqrt{4\pi n Z^2 e^2/M}$ is the ion plasma frequency,
and $\Gamma_{\rm p}\equiv Z^2 e^2/(a \omega_{\rm p})$. Then 
\begin{equation}
  \sigma_{\max} = n\omega_{\rm p}\left(\Gamma_{\rm p}\frac{\partial \zeta}{\partial\epsilon}
  +1.5\frac{\partial u_1}{\partial\epsilon}\right){\bigg\vert_{\epsilon_{\max}}} .
\end{equation}
The values of the first momentum and the Madelung constant at $\epsilon=0$ are $-0.904122$ and $0.47817$, while at $\epsilon=0.10895$ they are $-0.903061$ and $0.47940$. Thus, the scale of the $u_1$ change coincides with the scale of the $\zeta$ change, while for the solid neutron star crust $\Gamma_{\rm p} \gtrsim 100$. For instance, at $\kappa a = 4/7$ 
\begin{equation}
\frac{\partial \zeta}{\partial\epsilon}{\bigg\vert_{\epsilon_{\max}}} \approx 0.01938~, \qquad \frac{\partial u_1}{\partial\epsilon}{\bigg\vert_{\epsilon_{\max}}} \approx -0.012~.
\end{equation}
Consequently, the contribution of zero-point vibration energy to $\sigma_{\max}$ is a few percent or less and its influence on the breaking stress can be mostly neglected. For $T=0$ it is appropriate to write
\begin{equation}
  \sigma_{\max} = 0.01938 n \frac{Z^2e^{2}}{a}.
\end{equation}

At high temperatures ($T \gg T_{\rm p}$, where $T_{\rm p}\equiv\hbar \omega_{\rm p}$)
the total internal energy in the harmonic approximation is
\begin{equation}
U_{\rm M}+E_0+F \approx  NT\left[\Gamma\zeta+3 u_{\ln}-3\ln (T/T_{\rm p})\right]~, \label{f}    
\end{equation}
where $F\equiv 3NT\langle\ln (1-\exp (-\hbar \omega/T))\rangle$ is the thermal contribution, $u_{\ln}\equiv
\langle\ln (\omega/\omega_{\rm p})\rangle$. Then
\begin{equation}
\sigma_{\max}=nT\left(\Gamma\frac{\partial \zeta}{\partial\epsilon}+3\frac{\partial u_{\ln}}{\partial\epsilon}\right){\bigg\vert_{\epsilon_{\max}}}~.
\end{equation}

At $\kappa a = 4/7$ I obtain 
\begin{equation}
\sigma_{\max}=n\frac{Z^2 e^2}{a} \left(0.01938-\frac{0.9}{\Gamma}\right)~.
\label{temp}
\end{equation}
In this form the zero temperature and thermal contributions at $T \gg T_{\rm p}$ satisfactorily converges with the result of molecular dynamic simulations. The MD studies presented in \cite{CH10,CH12} determine $\epsilon_\text{max}$ and $\sigma_{\rm max}$ by direct simulations of the crystal evolution with increasing $\epsilon$. They use the same deformation as I do here and in \cite{KY20}.
In \cite{CH10} it was reported (it was obtained from approximation MD simulations) that at $\kappa a = 4/7$ the breaking stress is
\begin{equation}
\sigma_{\max}^{\rm MD}=n\frac{Z^2 e^2}{a} \left(0.0195-\frac{1.27}{\Gamma-71}\right)~.
\label{chu}
\end{equation}

Therefore, $\sigma_{\max}^{\rm MD}=0.0195nZ^2e^2/a$ at $T=0$ and the difference with my result is less than $1\%$. The slight difference in the results at $T \gg T_{\rm p}$  can be explained by the insufficient accuracy of our calculations (in contrast to  $u_1$, the $u_{\ln}$ moment requires more precise studies) and MD simulations (see, Figures 1-3 in \cite{CH12}). In any case, the agreement between results supports the statement that the complex-valued phonon modes control the crystal stability.

Note that with the accuracy in use, the difference between the thermal contribution at $\kappa a = 4/7$ and $\kappa a = 0$ is negligible, while the zero temperature contribution at $\kappa a = 0$ is $\sigma_{\max}=0.02007nZ^2e^2/a$. Thus, this gives a breaking stress
\begin{equation}
\sigma_{\max, 0}=n\frac{Z^2 e^2}{a} \left(0.02007-\frac{0.9}{\Gamma}\right)~.
\end{equation}

According to molecular dynamic simulations \cite{CH10} at $\kappa a = 4/7$ the thermal correction to $\sigma_{\max}$ can be neglected at $\Gamma \gtrsim 6.5 \times 10^3$ (changes become less than 1 $\%$, see Eq. (\ref{chu})) and at such high $\Gamma$ the zero temperature approach can be used. Whereas at $\Gamma \lesssim 6.5 \times 10^3$ the thermal correction to the breaking stress should be investigated together with the anharmonic corrections. 

It is also instructive to mansion about the influence of a magnetic field ($B$). At $T \gg T_{\rm p}$ the breaking stress depends on two parameters: $\zeta$ and $u_{\ln}$, but both parameters are independent on the magnetic field therefore at high temperatures $\sigma_{\max}$ is independent on $B$. At $T=0$ and at high magnetic field, the first momentum is determined only by the properties of the magnetic field and does not depend on the deformation, hence its contribution to $\sigma_{\max}$ is absent and the breaking stress remains the same as at $B=0$.

\section*{Acknowledgments}

\end{document}